\tolerance   10000
\magnification = 1200
\baselineskip=1.65\normalbaselineskip
\font\ti=cmti10  scaled 1000
\font\tibig=cmti12 scaled 1200
\font\bigrmtwenty = cmb10 scaled 2000 
\font\bigrmsixteen = cmb10 scaled 1000
\font\smallrm = cmr10 scaled 800
\font\smallrmb = cmb10 scaled 800
\input epsf

\def\headingfont{\tenrm}
\headline={\centerline {\headingfont \folio}}
\footline={\hfil}
\topskip=20pt
\let\origeqno=\eqno
\def\eqno(#1){\origeqno (\rm #1)}

\vskip  35pt
\noindent
{\centerline {\bigrmtwenty  Dynamical Friction and Resonance Trapping }}
\vskip  1pt
\noindent 
{\centerline {\bigrmtwenty   in Planetary Systems }}
\vskip   30pt
\noindent
{\centerline  {\bf{\tibig Nader Haghighipour $^\ast$}}}
\vskip  1pt
\noindent
{\centerline  {Department of Physics and Astronomy}}
\noindent
{\centerline { University of Missouri-Columbia}}
\noindent
{\centerline {Columbia ,  MO $\>$  65211 , USA}}
\vskip  20pt
\noindent
{\bigrmsixteen  ABSTRACT}
\vskip  2pt
\noindent
A restricted planar circular three-body system, consisting of the Sun 
and two planets, is studied as a simple model for a planetary system. 
The mass of the inner planet is considered to be larger and the system 
is assumed to be moving in a uniform interplanetary medium with constant 
density. Numerical integrations of this system indicate a resonance 
capture when the dynamical friction of the interplanetary medium is taken 
into account. As a result of this resonance trapping, the ratio of orbital 
periods of the two planets becomes nearly commensurate and the eccentricity 
and semimajor axis of the orbit of the outer planet and also its angular 
momentum and total energy become constant. It appears from the numerical 
work that the resulting commensurability and also the resonant values of 
the orbital elements of the outer planet are essentially independent of the 
initial relative positions of the two bodies. The results of numerical 
integrations of this system are presented and the first-order partially 
averaged equations are studied in order to elucidate the behavior of the 
system while captured in resonance.
\vskip  15pt
\noindent
{\bigrmsixteen  Key words:} planetary dynamics , resonance capture , 
                                                                averaging .
\vskip  25pt
\noindent
$^\ast$E-mail:nader@lula.physics.missouri.edu 
\vfill
\eject

\noindent
{\bigrmsixteen  1 $\>\>\>$  INTRODUCTION}
\noindent
\vskip  2pt

The dragging effect of the interplanetary medium and its role in formation and 
evolution of planetary systems are quite well known and have long been investigated by 
many authors such as Weidenschilling et al. (1985 $\&$ 1993) , Patterson 
(1987) , Peale (1993) , Malhotra (1993) ,  Beaug\'e et al. (1993 and 1994a$\&$b) , 
Murray (1994)  and Gomes (1995). Recently, it has also been shown that the
dynamical friction of this medium can account for the stability of planetary
orbits in the early stages of their dynamical evolution (Melita and Woolfson
1996, here after MW). Using a formulation developed by Dodd and McCrea (1952,
see also Binney and Tremaine 1987) and by integrating a general three-body system, 
MW showed that the system of Sun and two massive planets,
subject to dynamical friction and accretion, will be captured 
into a near mean-motion resonance when the inner body is more massive. As a 
result of this resonance trapping, the eccentricities of the both planets become 
constant and their semimajor axes decrease monotonically. 

A simple planetary model is studied here based upon the results of Melita 
and Woolfson (1996). The simplifications introduced in this study make it 
possible to analyze the averaged dynamics of the system, analytically. 
The model used in this paper is a restricted planar circular three-body 
system with an inner planet that is more massive. Since MW showed that 
the effect of accretion will not change the end results qualitatively, 
this effect is not included in the dynamical equations of the system here. 
However, the frictional effect of the interplanetary medium is the main 
source of non-gravitational dissipation and is therefore taken into account.

The model under investigation is presented in section 2. Section 3 is 
concerned with the results of the numerical integrations. In order to analyze 
the numerical results analytically, a newly developed averaging technique 
for dissipative systems is introduced in section 4. This averaging method 
has been developed by Chicone et al. (1996-7a$\&$b) and was used in their analysis 
of dynamical behavior of binary systems subject to external gravitational 
radiation as well as radiation reaction damping. This averaging 
technique provides a very helpful tool to study the dynamics of the system 
at resonance. In section 5, the  application of this method  
to the dynamical equations of the system are discussed. The first-order 
averaged system at resonance is presented in section 6 and in section 7, 
conclusions of this study are presented by reviewing the results and 
discussing the possible extensions and applications.

\vskip  30pt
\noindent
{\bigrmsixteen  2$\>\>\>$  THE MODEL}
\vskip  5pt
\noindent

A restricted planar circular three-body system consisting of a central star 
(hereafter $\cal S$) and two planets is considered with the inner planet
more massive. It is assumed that the gravitational attraction of the planets 
on the central star is so small that its motion is negligible. This allows 
for the consideration of an inertial coordinate system with its origin on 
$\cal S\>.$ It is also assumed that after taking all perturbative effects into 
account, the orbital motion of the inner planet (hereafter $\cal I$) is 
uniformly circular with a given period. Dynamics of the outer planet 
(hereafter $\cal O$) is thus the focus of attention and is determined by the 
gravitational attractions of the central star and the inner planet and also by 
dissipation due to dynamical friction caused by the interplanetary medium.

In the inertial coordinate system under consideration, the equation of motion 
for planet $\cal O$ can be written as 
\vskip  1pt
$$
{m_{_O}}\,{{{d^2}{\vec r}_{_O}}\over {dt^2}}\>=\>-\,{\cal G}\,{{Mm_{_O}}\over 
{r_{_O}^3}}\>{{\vec r}_{_O}}\>-\>{\cal G}\,{{{m_{_I}}{m_{_O}}}\over {|{{\vec r}
_{_O}}-{{\vec r}_{_I}}|^3}}\>({{\vec r}_{_O}}\,-\,{{\vec r}_{_I}})\>+\>
{{\overrightarrow R}_{f}}\>\>\>,
\eqno  (1)
$$
\vskip  1pt
\noindent
where the indices $I$ and $O$ stand for the inner and the outer planets   
respectively, and ${{\overrightarrow R}_f}$ represents the dynamical 
friction force of the interplanetary medium and is given by 
(Dodd $\&$ McCrea 1952, Binney $\&$ Tremaine 1987 )
$$
{{\overrightarrow R}_f}\>=\>-\>2\pi{\rho_{_0}}{{{{\cal G}^2}{m_{_O}^2}}
\over {W^3}}\,\ln \biggl(1\>+\>{{{S^2}{W^4}}\over {{{\cal G}^2}{m_{_O}^2}}}
\biggl)\,{\overrightarrow W}\>\>\>.
\eqno  (2)
$$
\noindent
In this equation, $\overrightarrow W$ is the relative velocity of $\cal O$
with respect to the medium, $\rho_{_0}$ is the uniform density of the medium 
and
$$
S\>=\>{r_{_O}}\,{\big({{m_{_O}}\over{2M}}\big)}^{1/ 3}\>\>\>,
\eqno  (3)
$$
\noindent
where $M$ represents the mass of the central star.

Equation (1) can be simplified by dividing both sides  by 
$m_{_O}$ and also by letting ${\overrightarrow R}={{\overrightarrow R}_f}/
{m_{_O}}$ and ${\vec r}\,=\,{{\vec r}_{_O}}\>.$ The main equation
under consideration can therefore be written as
$$
{{{d^2}{\vec r}}\over {dt^2}}\>=\>-\,{\cal G}\,{M\over {r^3}}\>{\vec r}
\>-\>{\cal G}\,{{m_{_I}}\over {|{\vec r}-{{\vec r}_{_I}}|^3}}\>
({\vec r}\,-\,{{\vec r}_{_I}}\,)\>+\>{\overrightarrow R}\>\>\>.
\eqno  (4)
$$
\vskip  5pt
\noindent
It is more convenient to write this equation in  dimensionless form. 
Introducing $t_{_0}$ and $r_{_0}$ as the quantities that carry units of time 
and length respectively, $t$ and $r$ can be written as $t\,=\,{t_{_0}}
\,{\hat t}$ and $r\,=\,{r_{_0}}\,{\hat r}\>,$  where $\hat t$ and $\hat r$ 
are their corresponding dimensionless variables. Equation (4) has now the form
$$
{{{d^2}{\vec {\hat r}}}\over{d{\hat t}^2}}\>=\>-\,{K}{{\vec {\hat r}}\over 
{{\hat r}^3}}\>-\> \varepsilon\,{K}\>{{({\vec {\hat r}}\,-\,
{{\vec {\hat r}}_{_I}})}\over
{|{\vec {\hat r}}\,-\,{{\vec {\hat r}}_{_I}}|^3}}\>+\>
{\overrightarrow {\hat R}}\>\>\>,
\eqno  (5)
$$
\noindent
where 
$$
{K}\>=\>{{{\cal G}M{t_{_0}^2}}\over {r_{_0}^3}}\>\>\>,
\eqno  (6)
$$
\vskip  1pt
$$\!\!\!\!\!
{\varepsilon}\>\>\>\>=\>\>\>{{m_{_I}}\over M}\>\>\>,
\eqno  (7)
$$
\vskip  1pt
\noindent
and
$$
{\overrightarrow{\hat R}}\>\>=\>{\overrightarrow R}\>{\big ({{t_{_0}^2}\over
{r_{_0}}}\big)}\>\>\>.
\eqno  (8)
$$
\vskip  1pt
\noindent

In the rest of the calculations, I choose a set of units such that 
$K\,=\,1\>.$ Therefore from equation (6), $t_{_0}$ and $r_{_0}$ will be 
related as ${{t_{_0}^2}\,=\,{{r_{_0}^3}/{{\cal G}M}}}\>.$ This relation 
implies that $r_{_0}$ can be viewed as the orbital radius of a planet that orbits 
$\cal S$  on a circular path with a period $2\pi{t_{_0}}\>.$ In the 
model presented in this paper, planet $\cal I$ has this type of motion. 
Therefore, in the rest of this paper, I set ${r_{_0}}\,=\,{r_{_I}}$ and
${t_{_0}}\,=\,{{T_{_I}}/ {2\pi}}$ where $T_{_I}$ represents the (given) orbital 
period of planet $\cal I\>.$ Dropping the hat signs, equation (5) can be 
simplified as
\vskip  10pt
$$
{{{d^2}{\vec r}}\over {d{t^2}}}\>=\>-{{\vec r}\over {r^3}}\,-\,
\varepsilon\>{{({\vec r}\,-\,{\bf{{\vec r}_{_I}}})}\over {|{\vec r}\,-\,
{\bf{{\vec r}_{_I}}}{|^3}}}\,+\, {\overrightarrow R}\>\>\>,
\eqno (9)
$$
\vskip  10pt
\noindent
where ${\bf{{\vec r}_{_I}}}$ is the {\ti unit} vector along 
${{\vec r}_{_I}}\>.$

Let us now assume that the motion is planar. Introducing a polar coordinate
system on the plane of the orbits and with its origin at the location of the 
central star, the dynamical equations of planet $\cal O$ will be given by
$$
{P_r}\>=\> {\dot r}\>\>\>,
\eqno  (10)
$$
\vskip  1pt
$$
{P_\theta}\>=\> {r^2}\,{\dot \theta}\>\>\>,
\eqno  (11)
$$
\vskip  1pt
$$
{{\dot P}_r}\>=\>{{P_\theta^2}\over {r^3}}\>-\>{1\over{r^2}}\>-\>
{{\varepsilon}\over {|{\vec r}-{\bf{{\vec r}_{_I}}}|^3}}\>
\big[r\,-\,\cos\,(\theta-{\theta_{_I}})\,\big]\>+\>
({R_x}\cos\theta\>+\>{R_y}\sin\theta)\>\>\>,
\eqno  (12)
$$
\noindent
\vskip  15pt
and
\vskip  1pt
$$
{{\dot P}_\theta}\,=\,-\,\varepsilon\,{{ r}\over {|{\vec r}-
{\bf{{\vec r}_{_I}}}|^3}}\,
\sin(\theta-{\theta_{_I}})\,+\,r(-\,{R_x}\sin\theta\,+\,{R_y}\cos\theta)\>\>\>,
\eqno  (13)
$$
\vskip  10pt
\noindent
where $R_x$ and $R_y$ are the $x$ and $y$ components of $\overrightarrow R\>,$ 
respectively, and ${\theta_{_I}} = {2\pi t}/{T_{_I}} \,+\,$ constant. In 
dimensionless form and with a time origin such that this constant value becomes 
zero, we have ${\theta_{_I}}\,=\,t\,,$ where $t$ is the dimensionless time variable.

In order to calculate $R_x$ and $R_y$, we need to turn our attention to the 
interplanetary medium. It is assumed that this medium is freely rotating 
around $\cal S$ (Kiang 1962 , Dormand $\&$ Woolfson 1974).
Therefore its velocity at any point is perpendicular to the position vector 
at that point with a dimensionless magnitude equal to ${v_\mu}={r^{-1/2}}\>.$
The relative velocity ${\overrightarrow W} \,=\, {\vec v} \,-\, 
{{\vec v}_\mu}$ , will therefore have a radial component equal to $\dot r$ 
and a tangential component equal to $r\,{\dot \theta}\,-\,r\,{\omega_\mu}$ 
where ${\omega_\mu}={r^{-3/2}}$ is the dimensionless angular frequency of the 
medium at distance $r\>.$ As a result, the magnitude of $\overrightarrow W$ 
will be equal to
\vskip  1pt
$$
{W^2}\>=\>{{\dot r}^2}\>+\>{r^2}{({\dot \theta}\>-\>{\omega_\mu})}^2\>\>\>. 
\eqno  (14)
$$
\vskip  1pt
\noindent
The components of $\overrightarrow W$ are easily obtained noting that at any
distance $r\>,$ the angle between ${\vec v}_\mu$ and the $x$-axis is equal to
$(\theta\,+\,{\pi\over2})\>.$  Therefore
\vskip  1pt
$$
{W_x}\>=\>{\dot r}\,\cos\theta\>\,-\,r\,({\dot \theta}\>-\>{\omega_\mu})\,
\sin\theta\>\>\>,
\eqno  (15)
$$
\vskip  1pt
\noindent
and
\vskip  1pt
$$
{W_y}\>=\>{\dot r}\,\sin\theta\>\,+\,r\,({\dot \theta}\>-\>{\omega_\mu})\,
\cos\theta\>\>\>.
\eqno  (16)
$$
\vskip  1pt
\noindent
From equation (2) and in dimensionless form,  $R_x$ and $R_y$ will then be equal to
\vskip  1pt
$$
{R_x}\>=\>-\> {A\over{W^3}}\>\ln \,(1+B{r^2}{W^4})\>\big[{\dot r}\cos\theta\,-\,
r({\dot \theta}-{\omega_\mu})\sin\theta\big]\>\>\>,
\eqno  (17)
$$
\vskip  1pt
\noindent
and
$$
{R_y}\>=\>-\> {A\over{W^3}}\>\ln \,(1+B{r^2}{W^4})\>\big[{\dot r}\sin\theta\,+\,
r({\dot \theta}-{\omega_\mu})\cos\theta\big]\>\>\>,
\eqno  (18)
$$
\vskip  8pt
\noindent
where $A$ and $B$ are positive parameters given by
\vskip  8pt
$$
A\,=\,2\,\pi\, \Bigl({{{\rho_{_0}}{r_{_I}^3}}\over M}\Bigr)\>
\Bigl({{m_{_O}}\over M}\Bigr)\qquad,
\qquad B\,=\,{2^{-\,{2/3}}}\>\Bigl({{m_{_O}}\over M}\Bigr)^{-\,{4/3}} \>\>\>,
\eqno  (19)
$$
\vskip  8pt
\noindent
such that $A<<1$ and $AB<<1$ for any realistic system.
\vskip  100pt
\noindent
{\bigrmsixteen  3 $\>\>\>$ NUMERICAL RESULTS}
\vskip 15pt
\noindent

Equations (10) to (13) were numerically integrated for different values 
of the perturbation parameter $\varepsilon$ and density $\rho_{_0}$ using a 
stiff integration routine. Following MW, integrations were first performed on the 
Sun-Jupiter-Saturn system starting from the present near (5:2) commensurability.
The density of the interplanetary medium was taken to be constant and equal to 
$10^{-11}\>$ Kg m$^{-3}$  which is equivalent to 16 times the mass of Jupiter 
uniformly spread in a sphere of radius 50 au. In complete agreement with 
the results of MW, a near (2:1) commensurability is obtained for all initial 
relative positions of the two planets. A typical case is presented in Figure 1 .
\vskip  25pt
\hskip  30pt
\epsfbox{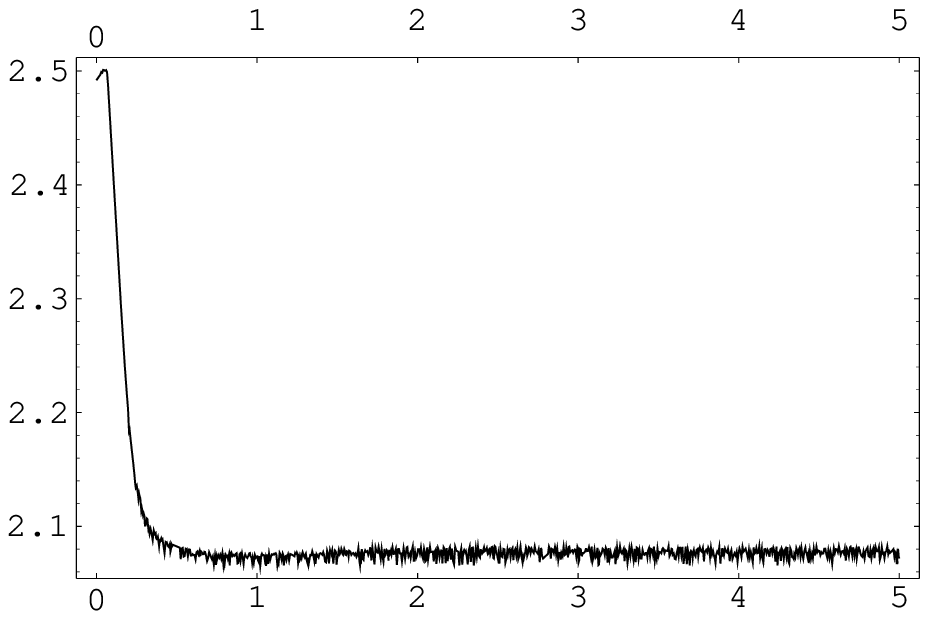}
\hfill
\vskip  10pt
\vbox{
\baselineskip=\normalbaselineskip
\smallrm
\noindent
{\smallrmb Figure 1.}  Graph of the ratio of orbital period of the outer planet 
(i.e., Saturn) to that of the inner one (i.e., Jupiter) against time. Integrations 
were performed with Jupiter on the  x-axis with $\theta_{_I}=0$ and Saturn at
$\theta_{_O}$=45$^\circ$. The initial values of $r\,,\,P_\theta$ and $P_r$ 
were calculated from $r=a(1-{e^2})/(1+e\cos \hat v)\,,\,
P_\theta^2 = a(1-{e^2})$ and $P_r = e\sin \hat v /{P_\theta}$ , where 
$a$=1.838046 is the present dimensionless value of the mean semimajor axis of 
Saturn's orbit, $e$ = 0.0556 is its present mean orbital eccentricity and 
$\hat v$ = 15$^\circ$ is its initial true anomaly (see section 4 and also 
Figure 4). The timescale is (10$^4$${T_{_I}}/{2\pi}$) years.}
\vskip  25pt
\noindent
As a result of this resonance trapping, the orbital eccentricity and 
semimajor axis of the outer planet (i.e., Saturn) and also its angular 
momentum and total energy become essentially constant (Figure 2). 
\vskip  10pt
$\!\!\!\!\!\!\!\!\!\!\!\!\!\!\!\!\!\!\!$
\epsfbox{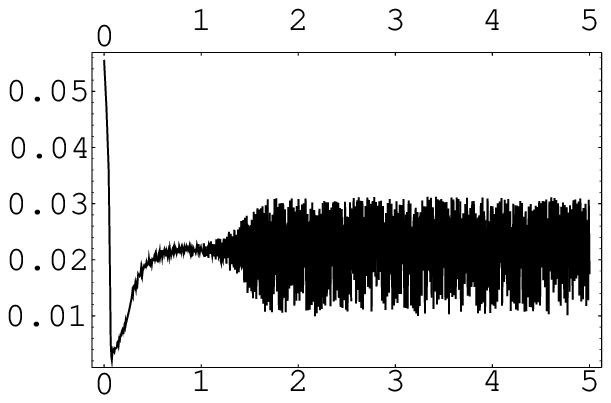}
\epsfbox{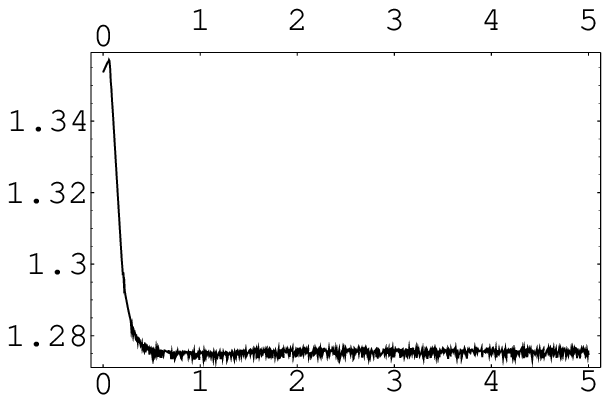}
\hfill
\vskip  5pt
$\!\!\!\!\!\!\!\!\!\!\!\!\!\!\!\!\!\!\!$
\epsfbox{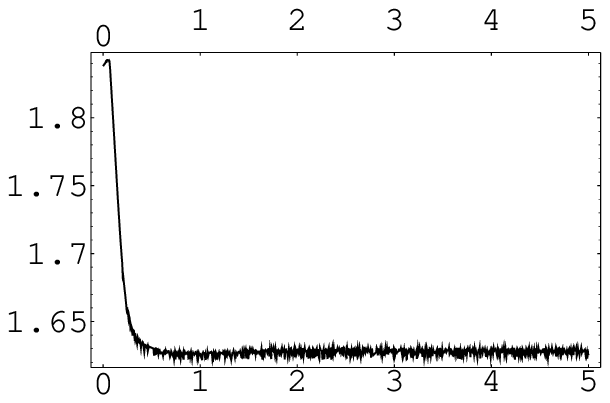}
\epsfbox{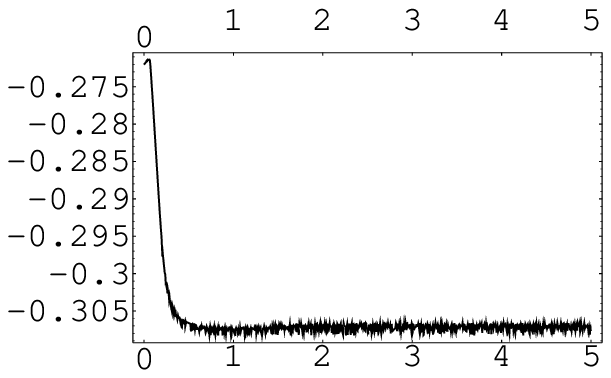}
\hfill
\vskip  1pt
\vbox{
\baselineskip=\normalbaselineskip
\smallrm 
\noindent
{\smallrmb Figure 2.} From top to bottom, orbital eccentricity and 
semimajor axis (left column) and angular momentum and energy 
(right column) of Saturn against time while captured in resonance. 
The initial conditions and the timescale are similar to those of Figure 1 .}
\vskip  20pt
\noindent  
Since the present density of the interplanetary medium is much smaller than 
10$^{-11}$Kgm$^{-3}$, integrations were also performed with lower values of 
density for the actual Sun-Jupiter-Saturn system. A typical case is shown in 
Figure 3 ,  where the system is captured in a (3:1) resonance but 
leaves this state after a short time.
\vskip  10pt
$\!\!\!\!\!\!\!\!\!\!\!\!\!\!\!$
\epsfbox{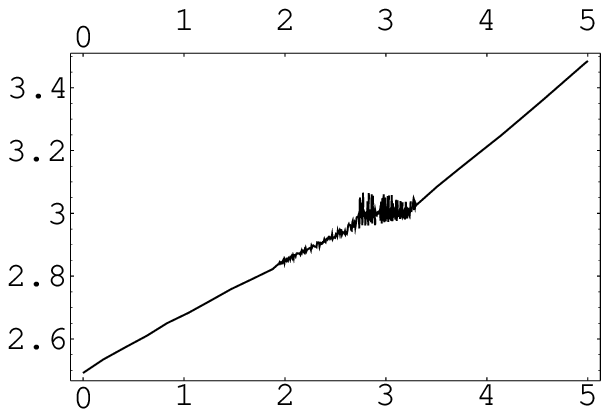}
\epsfbox{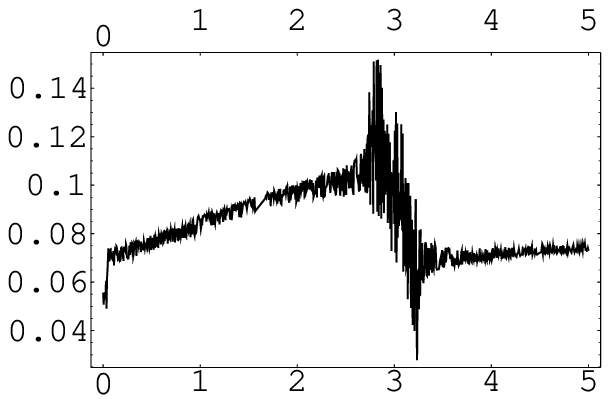}
\hfill
\vskip  10pt
\vbox{
\baselineskip=\normalbaselineskip
\smallrm 
\noindent
{\smallrmb Figure 3.} Graphs of the Ratio of orbital periods (left) and 
orbital eccentricity of Saturn (right) versus time for the actual 
Sun-Jupiter-Saturn system in an interplanetary medium with a 
density equal to 10$^{-14}\>$ Kg m$^{-3}$ . The system is captured  in 
a (3:1) resonance and leaves the resonance after a short time. 
The initial conditions of integration
are given by $(a,e,\theta,{\hat v})\,=\,$(1.838046 , 0.0556 , 75$^\circ$ , 
30$^\circ$) and the timescale is (10$^5$${T_{_I}}/{2\pi}$) years.}

In another series of runs, the planetary masses in the Sun-Jupiter-Saturn
system were changed while keeping their mass ratio constant and equal to
the present mass ratio of Jupiter and Saturn. The density of the 
interplanetary medium was kept equal to the 16 times the mass of the inner 
planet uniformly distributed inside a sphere of radius 50 au and the
mass of the central star was unchanged and equal to that of the Sun. 
Integrations were performed for different values of $\varepsilon$ 
and it was observed that only for the values of $\varepsilon$
in the range  ${10^{-2}}$ to ${10^{-4}}\,,$ the system was captured 
in resonance (Figure 4). The interesting result was that starting from present 
near (5:2) commensurability between Saturn and Jupiter, all resonances for
this range occurred near (2:1). It is important to mention that for the 
values of $\varepsilon$ smaller than $10^{-4}$, the planetary model
under consideration will no longer be physically valid. Numerical results
indicate that in these cases, the amplitude of oscillations of $r$ grows
rapidly with time which results in change of the configuration of the system.
That is, for long amplitudes, planet $\cal O$ approaches $\cal I$ and crosses
its orbit. This results in an exchange of positions between $\cal O$ and 
$\cal I$ and also causes the quantity $|{\vec r}-{\bf{{\vec r}_{_I}}}|$
in the dynamical equations of the system, reaches zero at certain times.
\vbox{
$\!\!\!\!\!\!\!\!\!\!\!\!\!\!\!\!\!\!$
\epsfbox{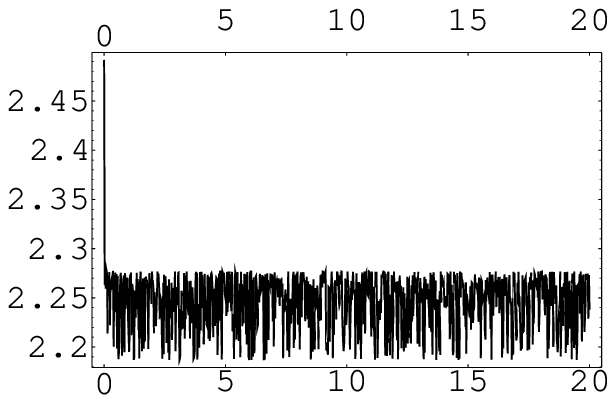}
\epsfbox{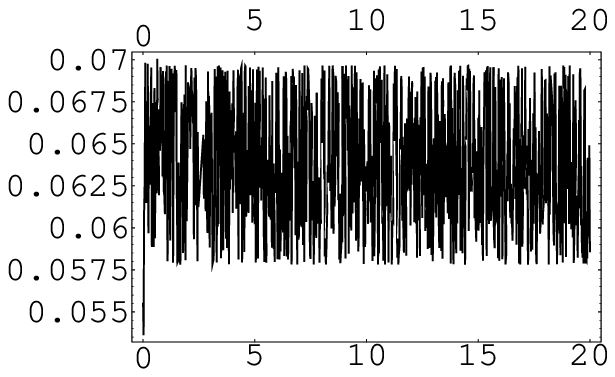}
\vskip  2pt
$\!\!\!\!\!\!\!\!\!\!\!\!\!\!\!\!\!\!$
\epsfbox{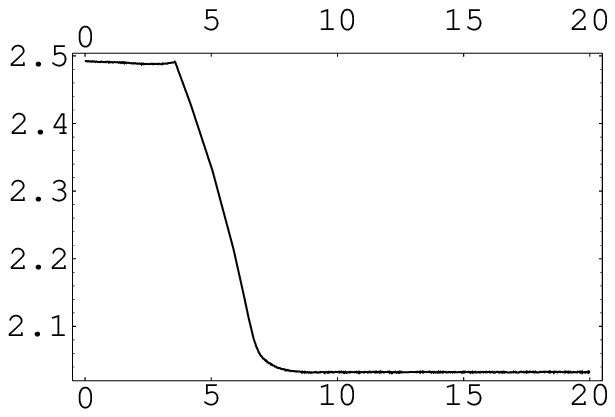}
\epsfbox{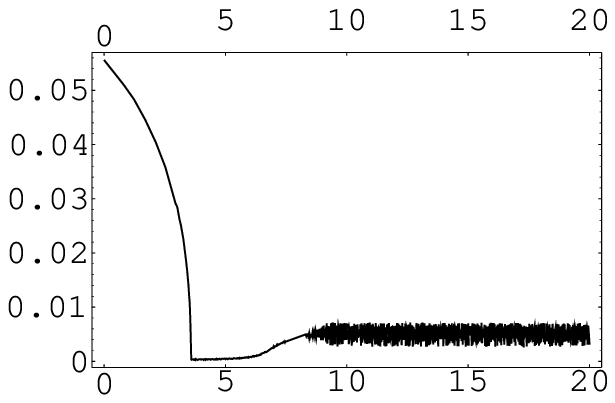}
\hfill
\vbox{
\baselineskip=\normalbaselineskip
\smallrm 
\noindent 
{\smallrmb Figure 4.} Graphs of the  ratios of orbital periods (left column) 
and orbital eccentricities (right column) versus time, for
$\varepsilon$ = 10$^{-2}$ (top), and $\varepsilon$ = 10$^{-4}$ (bottom). 
The initial values are given by $(a,e,\theta,\hat v)$ = (1.838046 , 
0.0556 ,75$^\circ$ , 30$^\circ$) and the timescale is 
(10$^4$${T_{_I}}/{2\pi}$) years.}
\vskip  10pt
\noindent

Numerical experiments on the actual Sun-Jupiter-Saturn system, were also
performed with larger values of the mass of the central star $M\,,$ but
no resonance capture was obtained. A typical case is presented in Figure 5.
\vskip 1pt 
$\!\!\!\!\!\!\!\!\!\!\!\!\!\!\!\!\!$ 
\epsfbox{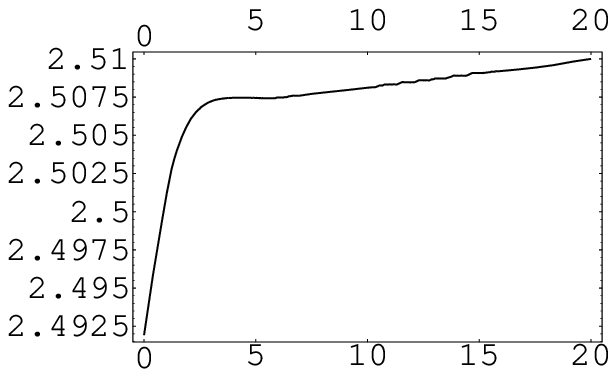}
\epsfbox{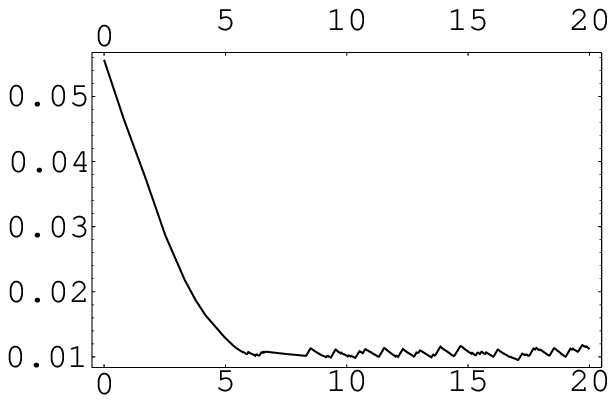} 
\hfill 
\vbox{ 
\baselineskip=\normalbaselineskip 
\smallrm 
\noindent
{\smallrmb Figure 5.} Ratio of orbital periods (left) and orbital 
eccentricity of the outer planet (right) against time for a central 
star with a mass equal to 1000 times the present mass of the Sun. 
The initial values and the timescale are the same as those in Figure 4.}}
\vskip  10pt
\noindent

The rest of this paper is devoted to the analysis of the results of 
the numerical integrations presented in this section. For this purpose, 
I will use a method of averaging called "Partial Averaging Near a 
Resonance" (Chicone et al. 1996) which is described in the next section.
\vskip  20pt
\noindent  
{\bigrmsixteen  4$\>\>\>$  AVERAGING}
\vskip  10pt
\noindent

Consider the following perturbation problem in a $k$-dimensional 
Euclidean space $\Re^k$ ,
$$
{\dot u}\,=\,f(u)\,+\,\epsilon b(u,t,\epsilon)\>\>\>,\>\>\>u
\in{\Re^k}\>\>\>.
\eqno  (20)
$$
\vskip  2pt
\noindent
\noindent 
In this system $\epsilon$ is the perturbation parameter, $b$ is a periodic 
function of time with a period $\eta>0$ and the overdot represents the 
derivative with respect to time $t\>.$ If the unperturbed system, i.e. 
${\dot u} \,=\,f(u)\>,u\in{\Re^k}$ is integrable, one can always 
find a set of canonical coordinate transformations to write equation (20) in 
terms of a set of action-angle variables 
$({\cal L},\phi)\,\in{\Re^k}\times{\aleph^\ell}$ such that  
$$
{\dot {\cal L}} \,=\, {\epsilon}\,{\cal F}({\cal L},\phi)\,+\,
O({\epsilon^2}) \>\>\>,
\eqno  (21)
$$
\noindent
and
$$
{\dot \phi} \,=\, \Omega({\cal L}) \,+\, {\epsilon}\,{\cal B}({\cal L},\phi)
\,+\,O({\epsilon^2})\>\>\>.
\eqno  (22)
$$
\vskip  2pt
\noindent
In these equations $\phi$ is a $2\pi$ modulo $\ell$-dimensional vector of 
angular variables and $\cal F$ and $\cal B$ are $2\pi$ periodic functions 
of $\phi\>.$ The most important feature of equations (21) and (22) is that 
in the angular equation (22), there appears a term which is independent of 
the perturbation parameter $\epsilon\>.$ This indicates that, in the first
order of approximations (i.e., $\epsilon$), $\phi(t)$ can 
be considered as a fast-changing angular variable while ${\cal L}(t)$ moves 
slowly away from its unperturbed value. Therefore, when studying the dynamics 
of the system (20), the time evolution of the averaged value of ${\cal L}(t)$ 
over $\phi(t)$ can be taken as a reasonable approximation to the evolution 
of ${\cal L}(t)$.

The above statement is often referred to as the "Averaging Principle." 
This principle states that if ${\cal L}\,=\,{{\cal L}_0}$ is the initial value 
for the action variable $\cal L$ in system (21) and (22), then for sufficiently 
small $\epsilon\>,$ the solution to the initial value problem
$$
{\dot J}\,=\,{\epsilon}\,{\bar{\cal F}}(J) \qquad ;\qquad J(0)\,=\,
{{\cal L}_0}\>\>\>,
\eqno  (23)
$$
\vskip  2pt
\noindent
will provide a useful approximation for the evolution of the action variable 
$\cal L$ with an error of order $\epsilon^2$ over a timescale of order 
$\epsilon^{-1}\>.$ In this equation 
$\bar {\cal F}$ is the averaged value of  $\cal F$ and is given by
$$
{\bar {\cal F}}\,(J)\,=\,{1\over {{(2\pi)}^\ell}}{\int_{\aleph^\ell}}
{\cal F}(J\,,\,\phi)\,d\phi\>\>\>.
\eqno  (24)
$$
\vskip 10pt
\noindent 

In the system presented in this paper, the unperturbed system (i.e. Sun-Saturn) 
is Hamiltonian. Also, in the perturbation problem given by equations (10) to (13),
$\varepsilon$ and $A$ are small quantities. Therefore we can use the 
averaging principle to analyze the dynamics of this system while captured in 
resonance. To this end, one needs to write the equations of motion in terms of 
appropriate action-angle variables that will provide us with a fast-changing 
angular variable. This is accomplished by writing the equations of motion in 
terms of Delaunay's variables which can be described as follows. 

If at any time $t$ all perturbative forces are removed, the orbit of 
the outer planet will become an ellipse with its focus always at the 
origin of coordinates. This ellipse is tangent to the actual orbit at point 
${\vec r}(t)$ and is therefore called an osculating ellipse 
(Brouwer $\&$ Clemence 1961, Kovalevsky 1967, Hagihara 1972 and 
Gutzwiller 1998). One can use the orbital elements of the osculating ellipse 
to introduce the Delaunay action-angle variables. In our planar problem, the 
relevant Delaunay variables are given by (Kovalevsky 1967, Sternberg 1969, 
Hagihara 1972)
$$
L\,=\,a^{1/2}\>\>\>,
\eqno  (25)
$$
$$
l\,=\,{\hat u}\,-\,e\sin{\hat u}\>\>\>,
\eqno  (26)
$$
$$
G\,=\,\bigl[{ a (1-e^2)}\bigr]^{1/2}\>\>\>,
\eqno  (27)
$$
\noindent
and 
$$
g\,=\,\theta\,-\,{\hat v}\>\>\>,
\eqno  (28)
$$
\vskip  2pt
\noindent 
where $a$ and $e$ are the semimajor axis and eccentricity of the osculating 
ellipse, respectively, and $l$ is the mean anomaly. In these 
equations $L$ and $G$ are action variables, while $l$ and $g$ are their 
corresponding angular variables. The variables $\hat v$ and $\hat u$ are, 
respectively, the true anomaly and the eccentric anomaly of the outer planet
as illustrated in Figure 6.

The polar coordinates and the momenta of the outer planet are related to the Delaunay
variables by ${P_\theta^2}\,=\,{G^2}\,=\,r(1+e\cos {\hat v})$ and ${P_r}\,=
\,{{e\,\sin {\hat v}}/G}\>.$ Using these equations along with replacing 
$\theta$ from equation (28), the equations of motion in terms of the Delaunay 
variables can be written as 
$$
{{dL}\over {dt}}\,=\,a\,(1-e^2)^{-{1/ 2}}\>\biggl[{F_r}\,e\,\sin{\hat v}\,+\,
{F_\theta}\,(1\,+\,e\,\cos{\hat v})\biggl]\>\>\>,
\eqno  (29)
$$
$$
{{dG}\over {dt}}\,=\,r\,{F_\theta}\>\>\>,
\eqno  (30)
$$
$$
{{dl}\over {dt}}\,=\,\omega\,+\,{r\over e}\,{a^{-{1/ 2}}}\,
\Bigl[{F_r}\bigl(-2e\,+\,\cos {\hat v}\,+\,e\,{\cos^2} {\hat v}\bigr)\,-\,
{F_\theta}\,\bigl(2\,+\,e\,\cos {\hat v}\bigr)\,\sin {\hat v}\Bigr]\>\>\>,
\eqno (31)
$$
\vskip  3pt
\noindent
and
\vskip  1pt
$$
{{dg}\over {dt}}\,=\,{1\over e}\,\Bigl[a(1-e^2)\Bigr]^{1/ 2}\,
\biggl[\,-{F_r}\cos {\hat v}\,+\,
{F_\theta}\Bigl(1\,+\,{1\over {1\,+\,e\,\cos{\hat v}}}\Bigl)
\sin{\hat v}\biggl]\>\>\>,
\eqno  (32)
$$
\noindent
where 
$$
{F_r}\>=\>\varepsilon\>{{\cos(\theta-{\theta_{_I}})\,-\,r}\over {|{\vec r}-
{\bf{\vec r_{_I}}}|^3}}\,+\,({R_x}\cos\theta\,+\,{R_y}\sin\theta)\>\>\>,
\eqno  (33)
$$
$$\!\!\!\!\!\!\!\!
{F_\theta}\>=\>-\,\varepsilon\,{{\sin(\theta-{\theta_{_I}})}\over {|{{\vec r}-
{\bf{\vec r_{_I}}}|^3}}}\,-\,({R_x}\sin\theta\,-\,{R_y}\cos\theta)\>\>\>,
\eqno  (34)
$$
\vskip  4pt
\noindent
and the Keplerian frequency of the osculating ellipse is given by
$\omega\,=\,L^{-3}\>.$
\hfill
\vskip  5pt
\epsfbox{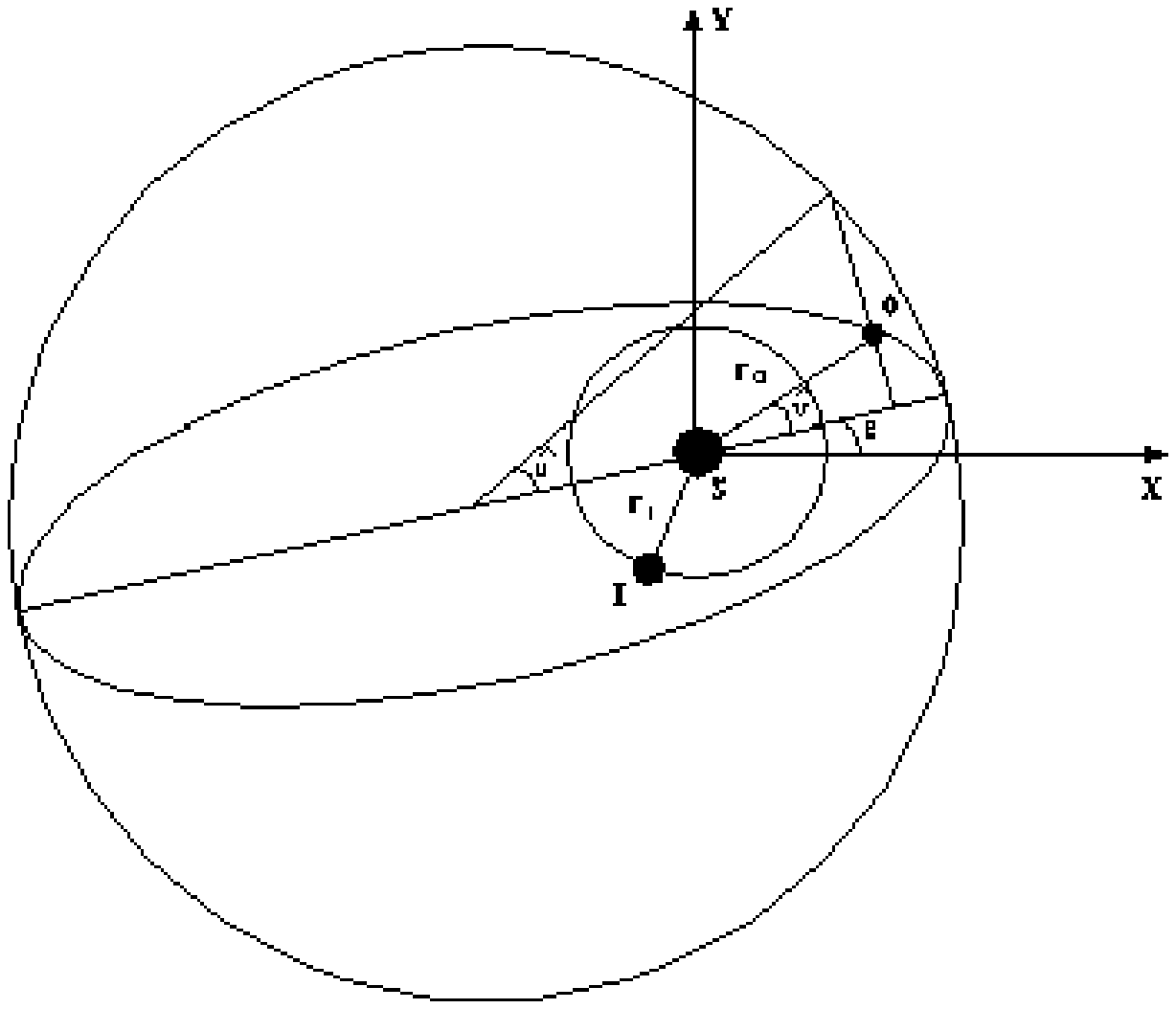}
\vskip  30pt
\vbox{
\baselineskip=\normalbaselineskip
\smallrm 
\noindent
{\smallrmb Figure 6.} The osculating ellipse for the outer planet. The central star 
is at one of the focal points. The orbit of the inner planet is a circle. 
The true anomaly $\hat v$ and the eccentric anomaly $\hat u$ are defined 
according to this figure.}
\vskip  30pt
Quantities $F_r$ and $F_\theta$ are, in fact, representing perturbations 
due to the  gravitational attraction of the inner planet and also the dynamical 
friction of the interplanetary medium. It is evident that 
only the angular variable $l\>,$ i.e. the mean anomaly, has an associated 
Keplerian frequency unaffected by perturbations. This immediately suggests 
that $l$ is the appropriate "fast" angular variable for averaging purposes. 
In the next section, the averaging principle is applied to the system of 
equations (29) to (32) and the averaged dynamics of the system while captured 
in resonance is discussed.

\vskip  20pt
\noindent
{\bigrmsixteen  5$\>\>\>$  AVERAGED SYSTEM AT RESONANCE}
\vskip  10pt
\noindent

As mentioned above, equations (29) to (32) represent the dynamics of the 
system in terms of action variables $(L,G)$ and angular variables $(l,g)\,.$
The objective of this section is to apply the averaging principle to the
equations (29) to (32) while the system is at resonance. 
In general, resonance occurs when the Keplerian frequency $\omega$ becomes 
commensurate with the frequency of the external perturbation. In the system 
presented in this paper, the external perturbation is due to gravitational 
attraction of the inner planet and its frequency is given by ${\omega_{_I}}=
2\pi/{T_{_I}}\>.$ Therefore in this system, resonance capture means that 
relatively prime integers $m$ and $n$ exist such that $m\omega=n\omega_{_I}\>.$ 
Since in this model the inner planet is orbiting the central star on a circular
path with a fixed orbital period, the immediate result of resonance trapping 
is a constant value for the orbital period of the outer planet. A constant 
orbital period in turn, results in a constant value for the semimajor axis 
of the osculating ellipse and therefore a constant value for the action 
variable $L\>.$ At resonance,  $L={L_0}={(m/n{\omega_{_I}})^{1/3}}\>.$ 
However, as it is evident from Figure 2, the value of $L\,=\,{a^{1/2}}$ 
during resonance is not entirely constant, but oscillates around its 
Keplerian value $L_0\>.$ Let $D$ be a measure of these oscillations and 
also let the oscillations of the mean anomaly from its Keplerian value 
${L_0^{-3}}\,t$ be given by $\varphi\>.$ Requiring $\dot L$ and $\dot l$ 
to have the same power of $\varepsilon$ in the lowest order of 
approximation, one can write 
$$
L\,=\,{L_0}\,+\, \varepsilon^{1/2}\,D\>\>\>,
\eqno  (35)
$$
\noindent
and
$$
l\,=\,{1\over {L_0^3}}\,t\,+\,\varphi\>\>\>,
\eqno  (36)
$$
where the first order of approximation here has perturbation parameter 
$\varepsilon^{1/2}\>.$
\noindent

Equations (35) and (36) are indeed the necessary transformations that one needs 
to employ in order to write equations (29) to (32) for the system at resonance.
In order to study the averaged dynamics of the system at this state, the averaging
method presented in section 4 must then be applied to the transformed equations.
However, it is important to mention that the averaging principle is valid only for
systems with one angular variable (Chicone et al. 1997a). In the system
presented in this paper, there are two angular variables $l$ and $g\,.$
One can, however, show that the average rate of variation of $g$ is in fact 
negligible and therefore it can be considered as a slow-changing quantity. 
This becomes evident by writing equations (29) to (32) as
$$
{{dL}\over {dt}}\>=\>-\,\varepsilon\,{{\partial {H_{ext}}}\over 
{\partial l}}\>+\>\varepsilon \, \Delta \,{{\cal R}_L}\>\>\>,
\eqno  (37)
$$
$$
{{dG}\over {dt}}\>=\>-\,\varepsilon\,
{{\partial {H_{ext}}}\over {\partial g}}
\>+\>\varepsilon\,\Delta\,{{\cal R}_G}\>\>\>,
\eqno  (38)
$$
$$\qquad
{{dl}\over {dt}}\>\>\>=\,{1\over{L^3}}\,+\,\varepsilon \,
{{\partial {H_{ext}}}\over {\partial L}}
\>+\>\varepsilon \, \Delta\,{{\cal R}_l}\>\>\>\>\>\>,
\eqno  (39)
$$
\noindent
and
$$
\!\!\!
{{dg}\over {dt}}\>=\,\varepsilon
{{\partial {H_{ext}}}\over {\partial G}}
\>+\>\varepsilon\,\Delta\,{{\cal R}_g}\>\>\>\>\>\>,
\eqno  (40)
$$
\vskip  2pt
\noindent
where $\Delta \,=\,AB/\varepsilon\>,$
\vskip  2pt
$$
{H_{ext}}\,=\,-\,{1\over {|{\vec r}\,-\,{\bf {{\vec r}_{_I}}}|}}\>\>\>,
\eqno  (41)
$$
\vskip  10pt
\noindent
and ${{\cal R}_L}\,,\,{{\cal R}_G}\,,\,{{\cal R}_l}$ and ${{\cal R}_g}$
represent the contributions of dynamical friction in their corresponding
equations. I will show in section 6 that from these contributions,
only ${{\cal R}_L}$ will appear in the first-order averaged system.

Equations (39) and (40) indicate that while $l$ has a perturbation-free
frequency $L^{-3}$, the rate of change of $g$ is proportional to
$\varepsilon\,.$ This implies that, in comparison to $\dot l$ , the rate 
of variation of $g$ can be
neglected and therefore, for the purpose of this study, equations (29)
to (32)can be considered to have only one fast-changing angular variable 
that is $l\,.$ Equations (37) and (39) have now the general form of the 
system (21) and (22) and can therefore be used as the appropriate grounds 
for applying our averaging method.
That means, after replacing $L$ and $l$ in equations of motion (37) to (40) 
by their equivalent expression given by (35) and (36) and averaging the resulting 
equations over the fast-changing angular variable $l$, the averaged equations 
will represent a system whose dynamics, in the first order of approximation, will
be the same as the dynamical behavior of the original 
system at resonance, for a time interval $\varepsilon^{-1/2}\,{t_{_0}}$ where 
${t_{_0}}$ is the unit of time introduced in section 2. 
To do this, let us differentiate equations (35) and (36) with respect 
to time and expand $1/ {L^3}$ in powers of $\varepsilon^{1/2}$ as
\vskip  1pt
$$
{1\over {L^3}}\>=\>{1\over {L_0^3}}\>\Biggl[1\>-\>{\varepsilon^{1/2}}\>
\Bigl({{3D}\over {L_0}}\Bigr)\>+\>{\varepsilon}\Bigl({{6D^2}\over {L_0^2}}
\Bigr)\>+\>O(\varepsilon^{3/ 2})\Biggr]\>\>\>.
\eqno  (42)
$$
\vskip  10pt
\noindent
Equations (37) to (40) will now  have the form 
$$\!\!\!\!\!\!\!\!\!
\!\!\!\!\!\!\!\!\!\!\!\!\!\!\!\!\!\!\!\!\!\!\!\!\!\!\!\!\!\!\!\!
{\dot D}\>=\>-\,{\varepsilon^{1/2}}\>{F_{11}}\>-\>
\varepsilon D\>{F_{12}}\>+\>O({\varepsilon^{3/ 2}})\>\>\>,
\eqno  (43)
$$
$$
\!\!\!\!\!\!\!\!\!\!\!\!\!\!\!\!\!\!\!\!\!\!\!\!\!\!\!\!\!\!\!\!\!\!\!
\!\!\!\!\!\!\!\!\!\!\!\!\!\!\!\!\!\!\!\!\!\!\!\!\!\!\!\!\!\!\!\!\!\!\!\!\!
\!\!\!\!\!
{\dot G}\>=\>-\,\varepsilon\>{F_{22}}\>+\>O({\varepsilon^{3/ 2}})\>\>\>,
\eqno (44)
$$
$$
\!\!\!\!\!\!\!
{\dot \varphi}\>\>=\>-\>{\varepsilon^{1/2}}\>\Bigl({{3D}\over {L_0^4}}\Bigr)
\>+\>\varepsilon\>\biggl({{6D^2}\over {{L_0}^5}}\>+\>F_{32}\biggr)\>+\>
O({\varepsilon^{3/ 2}})\>\>\>,
\eqno(45)
$$
$$
\!\!\!\!\!\!\!\!\!\!\!\!\!\!\!\!\!\!\!\!\!\!\!\!\!\!\!\!\!\!\!\!\!\!\!
\!\!\!\!\!\!\!\!\!\!\!\!\!\!\!\!\!\!\!\!\!\!\!\!\!\!\!\!\!\!\!\!\!\!\!\!\!
\!\!\!\!\!\!\!\!\!\!
{\dot g}\>\>=\>\varepsilon\>F_{42}\>+\>O({\varepsilon^{3/ 2}})\>\>\>,
\eqno  (46)
$$
\vskip  15pt
\noindent
where
$$
F_{11} \,=\, {\partial \over {\partial l}}\,H_{ext}\,-\,{\Delta}\,
{{\cal R}_L}\>\>\>,
\eqno  (47)
$$
$$
F_{22} \,=\, {\partial \over {\partial g}}\,H_{ext}\,-\,{\Delta}\,
{{\cal R}_G}\>\>\>,
\eqno  (48)
$$
$$
F_{32} \,=\, {\partial \over {\partial L}}\,H_{ext}\,+\,{\Delta}\,
{{\cal R}_l}\>\>\>,
\eqno  (49)
$$
$$
F_{42} \,=\, {\partial \over {\partial G}}\,H_{ext}\,+\,{\Delta}\,
{{\cal R}_g}\>\>\>,
\eqno  (50)
$$
\vskip  10pt
\noindent
and $F_{12}\,=\,{{\partial F_{11}}/ {\partial L}}$ are evaluated at
$({L_0}\,,\,G\,,\,{{L_0^{-3}}\,t}\,+\,\varphi\,,\,g)\>.$ In labeling these 
terms, I have followed the convention of Chicone et al. (1997b).

The averaged equations of the system at resonance are obtained by averaging 
equations (43) to (46) over the mean anomaly $l\,.$ However, application
of the averaging principle requires an appropriate averaging transformation
that renders the system of (43) to (46) into a new system such that in the
lowest order, it becomes exactly the first-order 
averaged system (Chicone et al. 1997b). To obtain this 
averaging transformation, one has to define 
\vskip 1pt
$$
\lambda\,(G,\varphi\,,g\,,t)\,=\,
{F_{11}}\,\big(G,{1\over {L_0^3}}\,t\,+\,\varphi\,,g\,,t\,\big)
\,-\,{{\bar F}_{11}}\>\>\>,
\eqno  (51)
$$
\noindent
where
\vskip  1pt
$$
{{\bar F}_{11}}(G,\varphi,g) \,=\,
{{\omega_{_I}}\over {2\pi m}}{\int_0^{2\pi m/{\omega_{_I}}}}{F_{11}}
(G,{{1\over{L_0^3}}\,t\>+\>\varphi},g,t)\,dt\>\>\>,
\eqno  (52)
$$
\vskip 8pt
\noindent
is the averaged value of $F_{11}$ . Introducing $\Lambda (G,\varphi,g,t)$ by 
$$
{\partial \over {\partial t}}\,\Lambda (G,\varphi,g,t)\,=
\,\lambda (G,\varphi,g,t)\>\>\>,
\eqno  (53)
$$
\vskip  3pt
\noindent
such that ${\bar \Lambda}=0,$ one can use the averaging transformations
$D=\widehat D\>-\>{\varepsilon^{1/2}}\>\Lambda(\widehat G\,,\,
\widehat\varphi\,,\,\widehat g\,,\,t),$
\hfill 
$G=\widehat G\,,\,\varphi=\widehat\varphi\,,\,g=\widehat g$
to write the equations (43) to (46) as
\vskip  1pt
$$\!\!\!\!\!\!\!\!\!\!\!\!\!\!\!\!\!\!
{\dot {\widehat D}}\,=\,-\,{\varepsilon^{1/2}}\,{{\bar F}_{11}}\,-\,
\varepsilon\,{\widehat D}\>\Bigl({F_{12}}\,+\,{3\over {L_0^4}}\>
{{\partial \Lambda}\over {\partial \varphi}}\Bigr)\,+\,O({\varepsilon^{3/2}})
\>\>\>,
\eqno (54)
$$
$$\!\!\!\!\!\!\!\!\!\!\!\!\!\!\!\!\!\!\!\!\!\!\!\!\!\!
\!\!\!\!\!\!\!\!\!\!\!\!\!\!\!\!\!\!\!\!\!\!\!\!\!\!
\!\!\!\!\!\!\!\!\!\!\!\!\!\!\!\!\!\!\!\!\!\!\!\!\!
\!\!\!\!\!\!\!\!\!\!\!\!\!\!\!
{\dot {\widehat G}}\,=\,-\,\varepsilon {F_{22}}\,+\,
O({\varepsilon^{3/2}})\>\>\>,
\eqno  (55)
$$
$$\!\!\!\!\!\!\!
{\dot {\widehat \varphi}}\,=\,-\,{\varepsilon ^{1/2}}\>
\Bigl({{3\widehat D}\over {L_0^4}}\Bigr)\,+\,\varepsilon\,
\Bigl({{6{\widehat D}^2}\over {L_0^5}}\,+\,
{F_{32}}\,+\,{{3\Lambda}\over {L_0^4}}\Bigr)\,+\,
O({\varepsilon^{3/2}})\>\>\>,
\eqno  (56)
$$
$$\!\!\!\!\!\!\!\!\!\!\!\!\!\!\!\!\!\!\!\!\!\!\!\!\!\!
\!\!\!\!\!\!\!\!\!\!\!\!\!\!\!\!\!\!\!\!\!\!\!\!\!\!
\!\!\!\!\!\!\!\!\!\!\!\!\!\!\!\!\!\!\!\!\!\!\!\!\!\!
\!\!\!\!\!\!\!\!\!\!\!\!\!\!\!\!\!\!\!\!
{\dot {\widehat g}}\,=\,\varepsilon\,{F_{42}}\,+\,
O({\varepsilon^{3/2}})\>\>\>.
\eqno  (57)
$$
\vskip  5pt
\noindent
Since equations (10) to (13) represent a perturbation problem of order 
$\varepsilon\,,$ I will keep the terms proportional to $\varepsilon$ 
in equations above and neglect the $O({\varepsilon^{3/2}})$
terms. The averaged dynamics of the system will then be given by
\vskip 5pt
$$\!\!\!\!\!\!\!\!\!\!\!\!\!\!\!\!\!\!\!\!\!\!\!\!\!\!
\!\!\!\!\!\!\!\!\!\!\!\!\!\!\!\!\!\!\!\!\!\!\!\!\!\!
\!\!\!\!\!\!\!\!\!\!\!\!\!\!\!\!\!\!\!\!\!\!\!\!\!
\!\!\!\!\!\!\!\!\!
{\dot {\widetilde D}}\,=\,-\,{\varepsilon^{1/2}}\,{{\bar F}_{11}}\,-\,
\varepsilon\,{\widetilde D}\>{{\bar F}_{12}}\>\>\>,
\eqno  (58)
$$
$$\!\!\!\!\!\!\!\!\!\!\!\!\!\!\!\!\!\!\!\!\!\!\!\!\!\!
\!\!\!\!\!\!\!\!\!\!\!\!\!\!\!\!\!\!\!\!\!\!\!\!\!\!
\!\!\!\!\!\!\!\!\!\!\!\!\!\!\!\!\!\!\!\!\!\!\!\!\!
\!\!\!\!\!\!\!\!\!\!\!\!\!\!\!\!\!\!\!\!\!\!\!\!\!\!\!\!\!\!
\!\!\!\!\!\!\!\!\!\!\!\!\!\!\!
{\dot {\widetilde G}}\,=\,-\,\varepsilon {{\bar F}_{22}}\>\>\>,
\eqno  (59)
$$
$$\!\!\!\!\!\!\!\!\!\!\!\!\!\!\!\!\!\!\!\!\!\!\!\!\!\!\!\!\!\!
\!\!\!\!\!\!\!\!\!\!\!\!\!\!\!\!\!\!\!\!\!\!\!\!\!
{\dot {\widetilde \varphi}}\,=\,-\,{\varepsilon ^{1/2}}\>
\Bigl({{3\widetilde D}\over {L_0^4}}\Bigr)\,+\,\varepsilon\,
\Bigl({{6{{\widetilde D}^2}}\over {L_0^5}}\,+\,
{{\bar F}_{32}}\Bigr)\>\>\>,
\eqno  (60)
$$
$$\!\!\!\!\!\!\!\!\!\!\!\!\!\!\!\!\!\!\!\!\!\!\!\!\!\!
\!\!\!\!\!\!\!\!\!\!\!\!\!\!\!\!\!\!\!\!\!\!\!\!\!\!
\!\!\!\!\!\!\!\!\!\!\!\!\!\!\!\!\!\!\!\!\!\!\!\!\!\!
\!\!\!\!\!\!\!\!\!\!\!\!\!\!\!\!\!\!\!\!
\!\!\!\!\!\!\!\!\!\!\!\!\!\!\!\!\!\!\!\!\!\!\!\!\!\!\!\!\!
{\dot {\widetilde g}}\,=\,\varepsilon\,{{\bar F}_{42}}\>\>\>.
\eqno  (61)
$$
\vskip  2pt
\noindent
where ${{\bar F}_{12}}\,,\,{{\bar F}_{22}}\,,\,{{\bar F}_{32}}\,,$
and ${{\bar F}_{42}}$ are calculated in the same fashion as
${{\bar F}_{11}}$. 

\vskip  20pt
\noindent
{\bigrmsixteen  6$\>\>\>$  FIRST ORDER AVERAGED DYNAMICS}
\vskip  10pt
\noindent

In the present work, only the dynamics of the first-order averaged
system at resonance will be investigated. In this approximation, the
first-order averaged equations have perturbation parameter 
$\varepsilon^{1/2}$ and are given by 
$$
{\dot {\widetilde D}}\>=\>-\,{\varepsilon^{1/2}}\>{{\bar F}_{11}}
\quad\quad,\quad\quad
{\dot {\widetilde\varphi}}\>\>=\>-\>{\varepsilon^{1/2}}\>
\Bigl({{3{\widetilde D}}\over {L_0^4}}\Bigr)\quad\quad,\quad\quad
{\dot {\widetilde G}}\>=\>
{\dot {\widetilde g}}\>\>=\>0\>\>\>,
\eqno  (62)
$$
\noindent
where the terms with perturbation parameter $\varepsilon$ and higher 
have been neglected.

Equation(62) is, in fact, a general representation of the first-order
averaged system at resonance. In this equation, it is $F_{11}$ that 
contains information about the physical properties of the system.
For the three-body system presented in this paper, $F_{11}$ is
given by equation (47) where 
$$
{{\cal R}_L}\,=\,-\,{a\over {A\,B}}\, (1-{e^2})^{-1/2}\>
\Bigl\{{R_x}\,{\cal C}\,-\,{R_y}\,{\cal D}\,\Bigr\}\>\>\>.
\eqno  (63)
$$
\noindent
In this equation
$$
{\cal C}(L,G,l,g)\,=\,\sin ({\hat v}\,+\,g)\,+\,e\,\sin g\>\>\>,
\eqno (64)
$$
\noindent
and
$$
{\cal D}(L,G,l,g)\,=\,\cos ({\hat v}\,+\,g)\,+\,e\,\cos g\>\>\>.
\eqno  (65)
$$
\noindent

Since for the actual system of Sun-Jupiter-Saturn, numerical integrations 
have indicated a resonance lock with a near (2:1) commensurability
(see section 3), one has to set $m=2\,,\,n=1$ and ${\omega_{_I}}=1\,,$
which is the angular frequency of Jupiter in dimensionless form,
when one calculates ${\bar F}_{11}$ using integration (52).
This integration requires $H_{ext}$ and ${\cal R}_L$ to be written
in terms of the mean anomaly $l\,.$ Appendices A and B contain details of
these calculations. There, it has been shown that in the lowest order in
eccentricity, ${\cal R}_L$ is proportional to $e^2$. That is, 
$$
{{\cal R}_L}\,=\,-\,{1\over 2}\,{a_0^2}\,{e^2}\,
{\Bigl(1\,-\,{3\over 4}\,{\cos^2}l\Bigr)^{1/2}}\cos l \>\>\>,
\eqno  (66)
$$
\vskip  1pt
\noindent
where ${a_0} \simeq 1.625$ represents the resonant value of the 
semimajor axis of Saturn (see Figure 2).
It is evident from this equation that the averaged value of
${{\cal R}_L}$ over the mean anomaly $l$ will become zero. The first 
non-vanishing term in the averaged value of ${{\cal R}_L}$ appears as 
a term proportional to $e^3$. Numerical value of this term is so small 
that its contribution in dynamics of the first-order averaged system 
becomes actually negligible (see Figure 2 ; the numerical value of the
orbital eccentricity of Saturn at resonance is in average about
0.022 with an amplitude of oscillation equal to 0.01). 

The appearance of the higher orders of eccentricity in equation (66) 
requires $H_{ext}$ to be expanded to higher orders in $e$. However, it
turns out that even in expanding $H_{ext}$ to the second order in
eccentricity, non-vanishing terms after averaging, are at most of order
$10^{-4}$. A comparison between these values and the order of 
magnitude of the term proportional  to $e$ in that expansion
(i.e. $10^{-2}$), indicates that the contributions of $O({e^2})$ terms 
are entirely negligible. Therefore, in the rest 
of these calculations, I will neglect the contribution of dynamical 
friction and will consider 
${F_{11}} \simeq {\partial {H_{ext}}/\partial l}$.
The external Hamiltonian $H_{ext}$ will also be expanded only to the
first order in eccentricity. Appendix A shows that in that order,
the only contribution of $H_{ext}$ in ${{\bar F}_{11}}$ appears as 
the term proportional to $\cos (2l+g-{\theta_{_I}})$. Denoting this 
term by ${\cal H}_{ext}$, 
$$
{{\cal H}_{ext}}\,=\,-\,{{e\sigma}\over {2{a_0^2}}}\,\cos (2l\,+\,g\,-\,
{\theta_{_I}})\>\>\>,
\eqno  (67)
$$
\noindent
where
$$
\sigma\,=\,{\sum_{h=0}^\infty}\>{\Biggl[{{\Gamma ({3\over 2}+h)}\over
{{a_0^h}\>h!\>\Gamma({3\over 2})}}\Biggr]^2} 
\Biggl\{\,1\,+\biggl({{2h+3}\over {h+1}}\biggr)\>
\biggl[\,1\,-\,{3\over {4{a_0^2}}}\>\biggl({{2h+5}
\over {h+2}}\biggr)\biggr]\Biggr\}\>\>\>.
\eqno  (68)
$$
\vskip  7pt
\noindent
From equation (52), ${\bar F}_{11}$ will now be equal to
$$\!
{{\bar F}_{11}}\,=\,{{e\,\sigma}\over {a_0^2}} \> 
\sin (2\varphi\,+\,g)\>\>\>,
\eqno  (69)
$$
\noindent
and the first order averaged system at resonance will simply be obtained 
by substituting for ${{\bar F}_{11}}$ in equation (62). That is,
$$\!
\dot D\,=\,-\,{{e\,\sigma}\over {a_0^2}} \>{\varepsilon^{1/2}}
\,\sin (2\varphi\,+\,g)\>\>\>,
\eqno (70)
$$
\noindent
and
$$\!\!\!\!\!\!\!\!\!\!\!\!\!\!\!\!\!\!\!\!\!\!\!\!
\!\!\!\!\!\!\!\!\!\!\!\!
\dot \varphi \>\,=\, -\,{\varepsilon^{1/2}}\>\biggl({{3D}\over {L_0^4}}\biggr)\>\>\>,
\eqno  (71)
$$
\vskip  10pt
\noindent
where $G$ and $g$ are constants, and the tildes have been dropped 
for the sake of simplicity. From these two equations it follows that
$$
\ddot D\,-\,{{6\,e\,\sigma\,\varepsilon}\over {a_0^4}}\>D\>
\cos (2\varphi\,+\,g)\>=\>0
\>\>\>,
\eqno (72)
$$
\noindent
and
$$\!\!\!\!\!\!\!\!
\ddot \varphi\,-\,{{3\,e\,\sigma\,\varepsilon}\over {a_0^4}}\>
\sin (2\varphi\,+\,g)\>=\>0
\>\>\>. 
\eqno  (73)
$$ 
\noindent

Equation (73) is the equation of a mathematical pendulum. This  equation can be 
attributed to a Hamiltonian  $H_\varphi$ in the form
$$
{H_\varphi}(\varphi\,,\,{p_{_\varphi}})\,=\,
{\varepsilon^{1/2}}\Big[{1\over 2}\,{p_{_\varphi}^2} \,+\, U(\varphi)\Big]\>\>\>,
\eqno  (74)
$$
\noindent
with a potential given by
$$
U(\varphi)\,=\,{{3\,e\,\sigma}\over {2\,{a_0^4}}}\,
\cos (2\varphi\,+\,g)\,+\, Const.
\eqno (75)
$$

Figure 7 shows the graph of $U(\varphi)$ against $\varphi\,.$ 
The periodic characteristic of $U(\varphi)$ 
indicates that for all values of $\varphi$ except for those 
corresponding to the maxima of $U(\varphi)\,,$ the averaged 
system will fall into one of the potential wells and oscillates 
around a stable point forever. The resonance lock is established 
in this manner. For those values of $\varphi$ where $U(\varphi)$ 
becomes maximum, the system will be in an unstable equilibrium. A 
slight deviation from this instability will result in a resonance 
capture.
\vskip  15pt
\hskip  45pt
\epsfbox{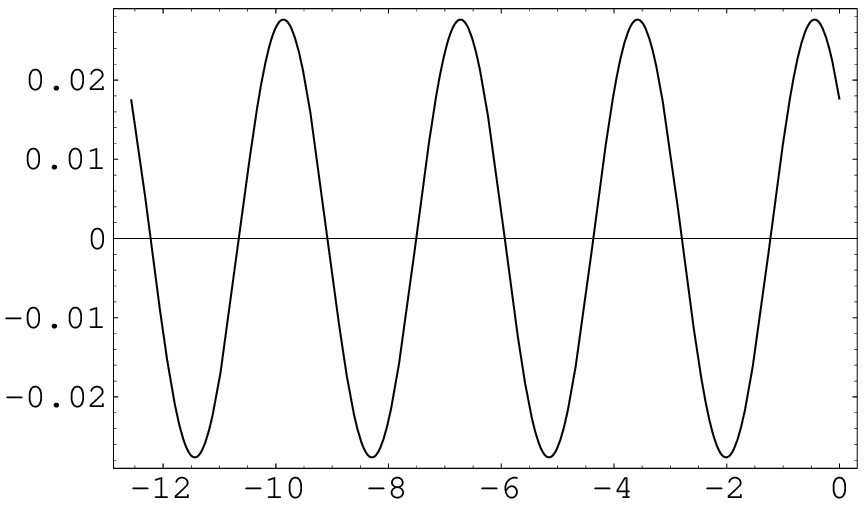}
\vskip  20pt
\vbox{
\baselineskip=\normalbaselineskip
\smallrm 
\noindent 
{\smallrmb Figure 7.} Graph of $U(\varphi)$ against $\varphi\,.$
The periodic nature of $U(\varphi)$ guarantees resonance capture for 
essentially all initial relative positions of the two planets.}
\vskip  3pt
\noindent

Equation (72), too, shows a harmonic behavior (Figure 8) . It is important to 
remember that $D$ represents the deviation of $L$ from its resonant value $L_0\>.$ 
From the averaging principle presented in section 4, it is expected that $D$ and 
$L$ will have the same general behavior for time intervals of duration  
${\varepsilon^{-{1/2}}}{t_{_0}}\>.$ This is illustrated in Figure 8 for the system 
presented in this paper .
\vskip  10pt
$\!\!\!\!\!\!\!\!\!\!\!\!\!\!\!\!\!\!$
\epsfbox{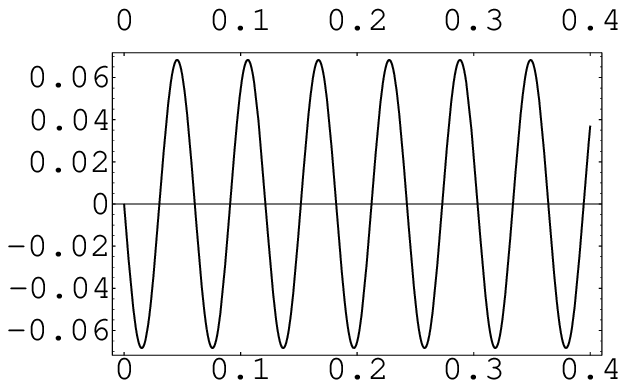}
\epsfbox{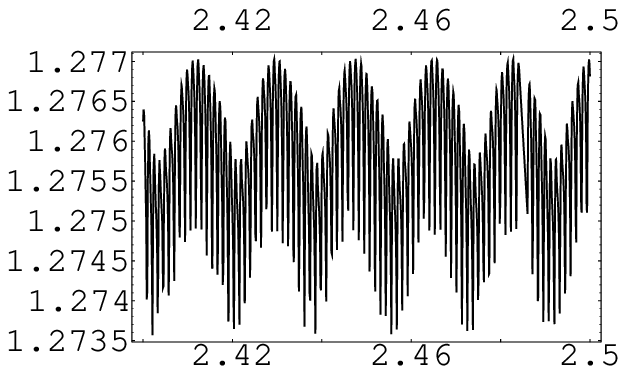}
\vskip  1pt
\vbox{
\baselineskip=\normalbaselineskip
\smallrm 
\noindent 
{\smallrmb Figure 8.} Graph of $D$ (left) and the action variable $L$ (right)agains time
while the system is at resonance. The original system shows a periodic behavior 
(right) which in a time interval ${\varepsilon^{-1/2}}{t_{_0}}$ is in  
very good agreement with the first-order averaged system (left). The timescale 
is (10$^4$${T_{_I}}/{2\pi}$) years. }
\vskip  15pt
\noindent
{\bigrmsixteen 7 $\>\>\>$ SUMMARY AND CONCLUSIONS}
\vskip  5pt
\noindent

In an attempt to study analytically, the resonance capture phenomenon reported by
Melita and Woolfson (1996), a restricted planar circular model of the Sun-Jupiter-
Saturn system subject to dynamical friction with a freely rotating homogeneous 
interplanetary medium was studied. Numerical integrations of this system indicated 
a resonance lock with the same commensurability as reported by MW. 
The method of partial averaging was employed to study the 
dynamical evolution of the system while captured in resonance. By averaging the 
equations of  motion over a fast-changing angular variable, these equations were 
reduced to the dynamical equations of a Hamiltonian system whose dynamics would be 
partially equivalent to the dynamical behavior of the main system. The application 
of this averaging method to the Sun-Jupiter-Saturn system where $\varepsilon=0.001\,,$ 
resulted in the first-order averaged system (70) and (71). Although this system is 
{\ti Hamiltonian} and it is not expected to fully illustrate the dynamical evolution 
of the main {\ti dissipative} system, but in a time interval 
${\varepsilon^{-1/2}}{T_{_I}}/{2\pi}\,,$ it can present a reasonable approximation to 
the dynamics of the original system. The harmonic characteristic of the potential 
function of this Hamiltonian system guarantees the occurrence of resonance lock 
for all initial relative positions of the two planets. Also, the time evolution of 
the action variable of the first-order averaged system illustrates an oscillatory 
behavior for the semimajor axis as well as the angular momentum of the main system 
over time scales of order $\varepsilon^{-1/2}$.

As mentioned above, the first-order averaged system at resonance is Hamiltonian. 
In general at this order, the dissipative effects of the non-gravitational 
perturbations appear as external torques  in the equation of the mathematical 
pendulum presented by the differential equation of the angular variable  $\varphi$. 
In the system presented in this paper, the averaged value of this torque in the lowest 
order in eccentricity, is proportional to $e^3$ and is entirely negligible when 
compared to the contribution of the gravitational attraction of 
the inner planet. The damping effect of this perturbation 
on the dynamical behavior of the original system can be illustrated by
extending the calculations to the second-order partially averaged system. This
will also allow for investigating the effects of the perturbation parameter
$\varepsilon$ and the density of the interplanetary medium $\rho_0$ on the time
variation of the orbital elements of the outer planet and their resonant values 
(Haghighipour, in preparation). 

The accreting force of the interplanetary medium was also neglected in this study.
Although the density of the interplanetary medium was considered much higher than
its present value, the masses of the planets were kept constant and equal to their
present values. Numerical integrations performed by Melita and Woolfson (1996) have
indicated that accretion will only change the results quantitatively. Therefore, 
one can apply the exact analysis presented in this paper, to the dynamical behavior 
of the system at resonance when accretion is also included.

Numerical integrations were also carried out for different values of the perturbation 
parameter $\varepsilon$ and the density of the interplanetary medium $\rho_{_0}$. 
The results indicated that when 
$\varepsilon$ is in the range 10$^{-2}$ to 10$^{-4}$, a Sun-Jupiter-Saturn-like 
system will be captured in a near (2:1) resonance when the density of
the interplanetary medium is taken to be equal to 16 times the mass of the inner
planet spread uniformly in a sphere of radius 50 au. However, for a real
Sun-Jupiter-Saturn system in a less dense medium, the resonance capture occurs
for a shorter time interval and also moves to (3:1) commensurability. 

The ideas presented in this paper can be applied to any dynamical system at resonance. 
Of particular interest would be extension of calculations to include the gravitational
tidal effects as another source of dissipation. This can be applicable to
analyze the dynamical behavior of triple systems (Arzoumanian 1996)
as well as newly discovered planetary systems (see for instance, Holland et al. 1998).
\vskip  15pt
\noindent
{\bigrmsixteen  ACKNOWLEDGEMENTS}
\vskip  10pt
\noindent

My deepest appreciation goes to Dr.B.Mashhoon for bringing the problem to my 
attention and also for critically reading the original manuscript and for 
fruitful discussions and suggestions. I am grateful to the Information and 
Access Technology Services of the University of Missouri-Columbia , especially 
to Dr.H.Tahani for providing me with supercomputing facilities without which 
numerical integrations of this project would have been impossible. And finally 
I would like to thank Dr.C. Chicone for stimulating discussions. 
\vskip  15pt
\noindent
{\bigrmsixteen REFERENCES}
\vskip 10pt 
\noindent 
Arzoumanian, Z., Joshi, K., Rasio, F.A.,  Thorsett, S.E. 1996, in Johnston S.,
Walker M.A., 
\vskip  1pt
\noindent
$\quad$Bailes M., eds, ASP Conf. Series Vol. 105, IAU 160-Pulsars: Problems 
and Progress, 
\vskip  1pt
\noindent
$\quad$p.525 
\vskip  1pt
\noindent
Beaug\'e C.,  Ferraz-Mello S., 1993, Icarus, 103, 301   
\vskip 1pt 
\noindent
Beaug\'e C., Ferraz-Mello S., 1994a, Icarus, 110, 239 
\vskip  1pt
\noindent
Beaug\'e C., Aarseth S.J. and Ferraz-Mello S., 1994b, MNRAS, 270, 21 
\vskip  1pt
\noindent
Binney J., Tremaine S., 1987, Galactic Dynamics, in Princeton
Series in Astrophysics, 
\hfill
\vskip  1pt
\noindent
$\quad$Princeton Univ. Press, Princeton, p.420
\vskip  1pt
\noindent
Brouwer D., Clemence G.M.,$\>$ 1961,$\>$ Methods of Celestial Mechanics, 
$\>$Academic Press, 
\hfill
\vskip  1pt
\noindent
$\quad$New York, p.541  
\vskip 1pt 
\noindent 
Chicone C., Mashhoon B., Retzloff D.G., 1996, Ann.Inst.
Henri Poincar\'e, Physique 
\hfill
\vskip  1pt
\noindent
$\quad$Th\'eorique, 64, 87  
\vskip 1pt 
\noindent 
Chicone C., Mashhoon B., Retzloff D.G., 1996, J.  Math.  Phys., 37, 3997  
\vskip 1pt 
\noindent 
Chicone C., Mashhoon B., Retzloff D.G., 1997a, CQG, 14, 699  
\vskip 1pt 
\noindent 
Chicone C., Mashhoon B., Retzloff D.G., 1997b, CQG, 14, 1831  
\vskip 1pt
\noindent 
Dodd K.N., McCrea W.H., 1952, MNRAS, 112, 205  
\vskip  1pt
\noindent
Dormand J.R., Woolfson M.M., 1974, Proc.Roy.Soc.Lond.A, 340, 349
\vskip  1pt
\noindent
Gomes R.S., 1995, Icarus, 115, 47 
\vskip 1pt 
\noindent 
Gutzwiller M.C., 1998, Rev.Mod.Phys., 70, 589   
\vskip 1pt 
\noindent 
Haghighipour N., 1998, in preparation
\vskip  1pt
\noindent
Hagihara Y., 1972, Celestial Mechanics, Vol 2, MIT Press,  
Cambridge, p.267 
\vskip  1pt
\noindent
Holland W.S., Greaves J.S., Zuckerman B., Webb R.A., McCarthy C., Coulson I.M.,
\vskip  1pt
\noindent
$\quad$Walther D.M., Dent W.R.F., Gear W.K., Robson I., 1998, Nature, 392, 788
\vskip  1pt
\noindent
Kiang T., 1962, MNRAS, 123, 359
\vskip  1pt
\noindent
Kovalevsky J., 1967, Introduction to Celestial Mechanics,  
Springer-Verlag, New York, p.38  
\vskip 1pt 
\noindent 
Malhotra R., 1993, Icarus, 106, 264  
\vskip 1pt 
\noindent 
Melita M.D., Woolfson M.M., 1996, MNRAS, 280, 854  
\vskip 1pt 
\noindent
Murray C.D., 1994, Icarus, 112, 465 
\vskip  1pt
\noindent
Patterson C.W., 1987, Icarus, 70, 319 
\vskip  1pt
\noindent
Peale S.J., 1993, Icarus, 106, 308 
\vskip 1pt
\noindent 
Sternberg S., 1969, Celestial Mechanics, Vol 1, Benjamin, New York, p.110   
\vskip 1pt 
\noindent
Weidenschilling S.J., Davis D.R., 1985, Icarus, 62, 16 
\vskip  1pt
\noindent
Weidenschilling S.J., Jackson A.A., 1993, Icarus, 104, 244 

\vskip  65pt
\noindent
{\bigrmsixteen  APPENDIX $\>$ A $\,:\,$ SIMPLIFICATION OF 
                                       $\bf{\mit{H_{ext}}}$}
\vskip  5pt

From equation (41) and in dimensionless form, $H_{ext}$ is given by
$$
H_{ext}\,=\,-\,{\bigl[{r^2}\,-\,2r\cos (\theta\,-\,{\theta_{_I}})\,+\,1\bigr]
^{-{1\over 2}}}\>\>\>.
\eqno  (A1)
$$
It is evident from this equation that in order to express $H_{ext}$ in terms 
of the mean anomaly $l\,,$ one needs to write $r$ and $\cos (\theta-{\theta_{_I}})$
in terms of $l\,.$ This is possible by writing $r={G^2}{(1+e\cos {\hat v})^{-1}}$
and $\theta=g+{\hat v}$, and expanding $\cos {\hat v}$ and $\sin {\hat v}$ as
(Kovalevsky 1967)
$$
\cos {\hat v}\,=\,-\,e\,+\,2\,\Bigl({{1-{e^2}}\over e}\Bigr)\,
{\sum_{j=1}^{\infty}}\,\cos (jl)\>{J_{_j}}(je)\>\>,
\eqno  (A2)
$$
\noindent
and
$$
\sin {\hat v}\,=\,(1-{e^2})^{1/2}\,{\sum_{j=1}^{\infty}}\,
\sin (jl)\>\bigl[{J_{_{j-1}}}(je)\,-\,{J_{_{j+1}}}(je)\bigr]\>\>.
\eqno (A3)
$$
\vskip  7pt
\noindent
In these equations, $J_{_j}$ is  Bessel function of order $j$ and is given by
$$
{J_{_j}}(s)\,=\,{\sum_{\nu=0}^\infty}\>{{(-1)^\nu}\over {\nu!\,(j+\nu)!}}\>
{\biggl({s\over 2}\biggr)^{j+2\nu}}\>\>\>.
\eqno  (A4)
$$
\vskip  7pt
\noindent
As mentioned in section 6, $H_{ext}$ is expanded to the first order in
eccentricity. Substituting for $\cos {\hat v}$ and $\sin {\hat v}$ in the
expressions for $r$ and $\cos (\theta-{\theta_{_I}})$ and considering terms
of order $e\,,$ one can therefore write
$$
r\,\simeq\,a\,(\,1\,-\,e\,\cos l)\>\>\>,
\eqno  (A5)
$$
\noindent
and
$$
\cos (\theta\,-\,{\theta_{_I}})\,\simeq\,\cos (l+g-{\theta_{_I}})\,+\,
e\,\Bigl[\cos (2l+g-{\theta_{_I}})\,-\,\cos (g-{\theta_{_I}})\,\Bigr]\>\>\>.
\eqno  (A6)
$$
\vskip  5pt
\noindent
The external Hamiltonian $H_{ext}$ will therefore be given by
$$\eqalign {\!\!\!\!\!\!
{H_{ext}}= - & {\bigl[1+{a^2} - 2a\cos (l+g-{\theta_{_I}})\bigr]^{-{1\over 2}}}\cr
&\qquad\qquad\qquad
\biggl[1+{1\over 2}\>e\>{{2{a^2}\cos l +a \cos (2l+g-{\theta_{_I}})-
3a\cos (g-{\theta_{_I}})}\over {1+{a^2}-2a\cos (l+g-{\theta_{_I}})}}\biggl]
\>\>\>.\cr}
\eqno  (A7)
$$
\vskip  7pt

At resonance, the semimajor axis of the outer planet is almost constant with an
average value ${a_0}\simeq 1.625\,.$ One can use this to simplify equation
(A7) by expanding $H_{ext}$ in terms of $a_0^{-1}$. 
The first term of this equation can be expanded using the relation 
\vskip 1pt
$$
{1\over {|{{\vec r}_{1}}\,-\,{{\vec r}_{2}}|}}\,=\,
{1\over {r_>}}\>{\sum_{N=0}^{\infty}}\>{\Bigl({{r_<}\over {r_>}}\Bigr)^N}\>
{P_{_N}}\,(\cos \Theta)\>\>\>,
\eqno  (A8)
$$
\vskip  8pt
\noindent
where $\Theta$ is the angle between ${\vec r}_{1}$ and ${\vec r}_{2}$ and 
${P_{_N}}\,(\cos \Theta)$ is Legender polynomial of order $N\>.$
In using equation (A8) in  (A7),
$\Theta=l+g-{\theta_{_I}}$. Therefore the Legender polynomial
${P_{_N}}\,(\cos \Theta)$ will produce terms proportional to 
$\cos N(l+g-{\theta_{_I}})$. Due to the harmonic nature of these terms, their 
averaged values obtained from equation (52) will all vanish. Therefore,
the term in equation (A7) that is independent of the eccentricity $e$  will 
have no contribution in ${\bar F}_{11}$ .
%
The remaining term in equation (A7)  can also be expanded using the relation
\vskip  1pt
$$
{\bigl(\,1\,-\,2\tau \cos \alpha\,+\,{\tau^2}\bigr)^{-\lambda}}\,=\,
{\sum_{q=0}^\infty}\>{C_q^\lambda}\,(\cos \alpha)\>{\tau^q}\qquad ,
\qquad |\tau|\,<\,1\>\>\>,
\eqno  (A9)
$$
\vskip  5pt
\noindent
where ${C_q^\lambda}(\cos \alpha)$ are Gegenbauer polynomials given by
\vskip  2pt
$$
{C_q^\lambda}\,(\cos \alpha)\,=\,{\sum_{h=0}^q}\>
{{\Gamma (\lambda+h)\>\Gamma (\lambda+q-h)}\over
{h!\>(q-h)!\>{[\Gamma (\lambda)]^2}}}\> \cos\big[(q-2h)\,\alpha\,\big]\>\>\>.
\eqno  (A10)
$$
\vskip  5pt
\noindent
In expanding the second term in equation (A7), $\lambda = 3/2\,,\,\tau={a_0^{-1}}$ 
and $\alpha = l+g-{\theta_{_I}}$ . From equation (36) where ${L_0^3}=2\>,$
the only non-vanishing terms in integration (52) are the ones proportional
to $\cos (2l+g-{\theta_{_I}})\>.$ Expansion (A10) produces terms proportional
to $\cos[(q-2h)\,(l+g-{\theta_{_I}})]\>.$ Therefore after multiplying the
numerator of the relevant term in equation (A7) by expansion (A9), the 
contribution of $\cos l$ appears in the terms where $q=2h+1\,,$ that of
$\cos (2l+g-{\theta_{_I}})$ in the terms where $q=2h$ and for
$\cos (g-{\theta_{_I}})$ when $q=2h+2\,.$ The contributing terms in
$H_{ext}$ (i.e., the terms with non-vanishing averaged values) can thus be
written as in equation (67).
\vskip  30pt
\noindent
{\bigrmsixteen  APPENDIX $\>$ B $\,:\,$ SIMPLIFICATION OF 
                                        $\bf{{\cal R}_{\mit L}}$}
\vskip  20pt
\noindent

Substituting for $R_x$ and $R_y$ from equations (17) and (18) in (63), 
${\cal R}_L$ can be written as
$$
{{\cal R}_L}\,=\,-\,{a\over {B\,{W^3}}}\,(1-{e^2})^{-1/2}\,
\ln (1\,+\,B\,{r^2}\,{W^4})
\Bigl [ \,a(1-{e^2})\,({\dot \theta}\,-\,{\omega_\mu})\,+
\,e\,{\dot r}\,\sin {\hat v}\Bigr]\>\>\>\>.
\eqno  (B1)
$$
\vskip  7pt
\noindent

The first step in simplifying ${\cal R}_L$ is to calculate the expression
inside the bracket of equation (B1). From equation (A5),
the angular frequency of the medium ${\omega_\mu}={r^{-3/2}}$ ,  and  
the angular frequency of the outer planet $\dot \theta=G{r^{-2}}\,,$   
are approximately equal to
\vskip  1pt
$$
{\omega_\mu}\,\simeq\,{a^{-3/2}}\,\Bigl(\,1\,+\,{3\over 2}\,e\,\cos l\,
\Bigr)\>\>\>,
\eqno  (B2)
$$
\noindent
and
\vskip  1pt
$$
{\dot \theta}\,\simeq\,{a^{-3/2}}\>\Bigl(\,1\,+\,2e\cos l\,
\Bigr)\>\>\>,
\eqno  (B3)
$$
\noindent
where $G$ has been replaced by its equivalent expression from equation (27).
Also from equation (A3) and to the first order in eccentricity, the 
radial momentum $\dot r\,=\,e{\sin \hat v}/G$ can be written as
$$
{\dot r}\,\simeq\,e\,{a^{-1/2}}\>{\sin l}\>\>\>.
\eqno  (B4)
$$
\vskip  10pt
\noindent
Substituting these values in the expression inside the bracket of 
equation (B1) and neglecting the terms of the second order and higher 
in eccentricity,
\vskip  1pt
$$
\Bigl [ \,a\,(1-{e^2})\,({\dot \theta}\,-\,{\omega_\mu})\,
+\,e\,{\dot r}\,\sin {\hat v}\Bigr]\,\simeq
{1\over 2}\,e\,{a^{-1/2}}\,\cos l\>\>\>.
\eqno  (B5)
$$
\vskip 10pt

It is now necessary to simplify the logarithmic term in 
equation (B1). Numerical computations indicate 
that, when the system is at resonance (i.e. 10$^{-4}$ $\leq$ $\varepsilon$ $\leq$ 
10$^{-2}$) , $B{r^2}{W^4}$ is very small in comparison to 1 (Figure B1). 
Therefore one can use the approximation $\ln (1\,+\,\delta)\,\simeq \, 
\delta $ for $\delta << 1$ to write this term as 
$\ln(1+B{r^2}{W^4})\cong B{r^2}{W^4}\>.$
\vskip  20pt
\hskip  70pt
\epsfbox{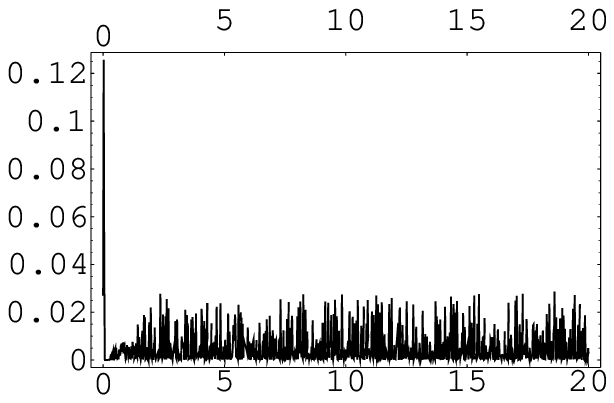}
\vskip  5pt
\vbox{
\baselineskip=\normalbaselineskip
\smallrm
\noindent
{\smallrmb Figure B1.} The graph of $B{r^2}{W^4}$ against time while the system 
is captured in resonance. $a,e,\theta,\hat v$ and $\varepsilon$ are respectively
equal to 1.838046 , 0.0556 , 45$^\circ$ , 15$^\circ$ and 0.001 . This figure 
shows that at resonance, $B{r^2}{W^4}$ is almost constant with an order of 
magnitude at most equal to 10$^{-2}\>.$  
The timescale is (10$^4$${T_{_I}}/{2\pi}$) years.}
\vskip  30pt
\noindent
Replacing $W^2$ from equation (14) and using equations (A5) and (B2) to (B4), $W$ 
can be written as
$$
W\,\simeq\,{a^{-1/2}}\,e\,\bigl(1\,-\,{3\over 4}\,{\cos ^2} l\Bigr)^{1/2}\>\>\>.
\eqno  (B6)
$$
\noindent
Finally from equations (B5) and (B6), ${\cal R}_L$ can be written 
as in equation (66) where in the lowest order, eccentricity appears 
as $e^2$.

\bye